\RequirePackage{lineno}
\documentclass[aps,prl,twocolumn,superscriptaddress,showpacs,preprintnumbers]{revtex4-2}
\usepackage[utf8]{inputenc}
\usepackage[dvipsnames]{xcolor}
\usepackage[normalem]{ulem}
\usepackage{comment}
\usepackage{overpic}
\usepackage{multirow}
\usepackage{booktabs} 
\usepackage{tikz}

\graphicspath{{./figures/}}
\usepackage[colorlinks,linkcolor=red,anchorcolor=blue,citecolor=red,breaklinks=true]{hyperref}
\usepackage{cleveref}
\newsavebox{\tablebox}
\usepackage{orcidlink}

\input{alias.tex}
\def\DpCS {{D^{+}\to K^{-}K^{+}\pi^{+}\pi^{0}}}
\def\DpCF {{D^{+}\to K^{-}\pi^{+}\pi^{+}\pi^{0}}}
\def\DsCF {{D_s^{+}\to K^{-}K^{+}\pi^{+}\pi^{0}}}
\def\DpDCS {{D^{+}\to K^{+}\pi^{-}\pi^{+}\pi^{0}}}
\def\DsCS {{D_s^{+}\to K^{+}\pi^{-}\pi^{+}\pi^{0}}}
\def\DpCSb {{D^{-}\to K^{+}K^{-}\pi^{-}\pi^{0}}}
\def\DpCFb {{D^{-}\to K^{+}\pi^{-}\pi^{-}\pi^{0}}}
\def\DsCFb {{D_s^{-}\to K^{+}K^{-}\pi^{-}\pi^{0}}}
\def\DpDCSb {{D^{-}\to K^{-}\pi^{+}\pi^{-}\pi^{0}}}
\def\DsCSb {{D_s^{-}\to K^{-}\pi^{+}\pi^{-}\pi^{0}}}

\def\simge{\mathrel{%
   \rlap{\raise 0.511ex \hbox{$>$}}{\lower 0.511ex \hbox{$\sim$}}}}
\def\simle{\mathrel{
   \rlap{\raise 0.511ex \hbox{$<$}}{\lower 0.511ex \hbox{$\sim$}}}}

\def\Acp {\mathcal{A}_{\CP}}
\def\ATodd {a^{T\text{-odd}}_{\CP}}
\def\DpDsp {D_{(s)}^{+}} 

\begin{document}

\title{\boldmath Search for $C\!P$ violation using $T$-odd correlations in 
$D_{(s)}^{+}\to K^{+} K^{-}\pi^{+}\pi^{0}$, $D_{(s)}^{+}\to K^{+} \pi^{-}\pi^{+}\pi^{0}$, and 
$D^{+}\to K^{-}\pi^{+}\pi^{+}\pi^{0}$ decays}

\begin{abstract}
We search for $\CP$ violation using $T$-odd correlations in five $D_{(s)}^{+}$ and $D_{(s)}^{-}$ four-body decays.
Our analysis is based on 980 $\rm fb^{-1}$ of data collected by the Belle detector at the KEKB energy-asymmetric 
$e^+e^-$ collider.
Our results for the $T$-odd \CP-violating parameter $\ATodd$ are:
$\ATodd(\DpCS) = (+2.6\pm 6.6\pm 1.3 )\times10^{-3}$,
$\ATodd(\DpDCS) = (-1.3\pm 4.2\pm 0.1 )\times10^{-2}$,
$\ATodd(\DpCF) = (+0.2\pm 1.5\pm 0.8 )\times10^{-3}$,
$\ATodd(\DsCS) = (-1.1\pm 2.2\pm 0.1 )\times10^{-2}$, and 
$\ATodd(\DsCF) = (+2.2\pm 3.3\pm 4.3 )\times10^{-3}$, 
where the uncertainties are statistical and systematic, respectively.
These results are the first such measurements and are all consistent with zero.
They include the first measurement for a $D^+_s$ singly Cabibbo-suppressed decay,
and the first measurement for a 
$D$ meson doubly Cabibbo-suppressed decay. We also measure $\ATodd$ in different subregions 
of phase space, where the decays are dominated by different intermediate resonance 
states such as 
$D^+\to\phi\rho^+$, $\Kstarzb\Kstarp$, and $\Kstarzb\rho^+$; and
$D_s^+\to \Kstarp\rho^{0}$, $\Kstarz\rho^{+}$, $\phi\rho^+$, and $\Kstarzb\Kstarp$.
No evidence for $\CP$ violation is found.
\end{abstract}

\noaffiliation
  \author{L.~K.~Li\,\orcidlink{0000-0002-7366-1307}} 
  \author{A.~J.~Schwartz\,\orcidlink{0000-0002-7310-1983}} 
  \author{E.~Won\,\orcidlink{0000-0002-4245-7442}} 
  \author{K.~Kinoshita\,\orcidlink{0000-0001-7175-4182}} 

  \author{I.~Adachi\,\orcidlink{0000-0003-2287-0173}} 
  \author{H.~Aihara\,\orcidlink{0000-0002-1907-5964}} 
  \author{S.~Al~Said\,\orcidlink{0000-0002-4895-3869}} 
  \author{D.~M.~Asner\,\orcidlink{0000-0002-1586-5790}} 
  \author{V.~Aulchenko\,\orcidlink{0000-0002-5394-4406}} 
  \author{T.~Aushev\,\orcidlink{0000-0002-6347-7055}} 
% ...
\collaboration{The Belle Collaboration}


\maketitle

\section{Introduction}

$\CP$ violation is required to explain the matter-antimatter asymmetry of the 
Universe~\cite{Sakharov:1967dj}. In the Standard Model~(SM), $\CP$ violation 
arises naturally due to the presence of a complex phase in the 
Cabibbo-Kobayashi-Maskawa~(CKM) matrix~\cite{Cabibbo:1963yz, Kobayashi:1973fv}. 
However, this source is insufficient to explain the absence of antimatter in 
the Universe. Thus it is important to search for new sources of $\CP$ violation
beyond the~SM. $\CP$ violation in charm decays is predicted to 
be very small, $\simle 10^{-3}$~\cite{Brod:2011re,Li:2012cfa,Cheng:2012wr}, 
and observing it significantly above this level 
would imply physics beyond the SM~\cite{PhysRevD.88.074011}. 
Charm decays proceeding via singly Cabibbo-suppressed~(SCS)
amplitudes are especially promising for such searches, as 
these amplitudes can have internal loops containing new 
mediators or couplings~\cite{Brod:2011re}.
To date, the only observation of $\CP$ violation in charm decays has been reported in 
the SCS decays $D^0\to\Kp\Km$ and $D^0\to\pip\pim$~\cite{LHCb:2019hro}. 
In this paper, we present the first search for $\CP$ violation 
in several additional Cabibbo-suppressed decay channels. 

We study four-body decays $\DpDsp\to P_1 P_2 P_3 P_4$,
where $P$ corresponds to a pseudoscalar meson,
by measuring the triple product of the momenta of three of the four 
final-state particles: 
$C_T\equiv \mbox{\boldmath $\vec{p}$}_1\cdot(\mbox{\boldmath{$\vec{p}$}}_2\times\mbox{\boldmath $\vec{p}$}_3)$. 
These momenta are evaluated in the $\DpDsp$ rest frame. 
We define an analogous observable $\CTbar$ for charge-conjugate 
$D^-_{(s)}$ decays. These observables are $T$ odd,
i.e., they change sign under time-reversal. 
Thus, if the distribution of $C_T$ is
asymmetric about zero, this could indicate $T$ violation. 
To quantify such an asymmetry, we define the variables
\begin{eqnarray}
A_{T} & \equiv & \frac{ N(C_T>0) - N(C_T<0) }{ N(C_T>0) + N(C_T<0) }\,, \\
\ATbar & \equiv & \frac{ \overline{N}(-\CTbar>0) - \overline{N}(-\CTbar<0) }
                   { \overline{N}(-\CTbar>0) + \overline{N}(-\CTbar<0) }\,, 
\label{eqn:AT}
\end{eqnarray}
where $N$ and $\overline{N}$ are the yields of $D_{(s)}^+$ and $D_{(s)}^-$ decays, respectively. 
The minus signs included in the definition of $\ATbar$ correspond to a parity transformation; 
in this manner $A_T$ and $\ATbar$ are \CP-conjugate quantities.
However, 
final-state 
interactions~(FSI) can also give rise to nonzero $A_T$ 
or $\ATbar$ and thus mimic $T$ violation. To isolate $T$ violation, we measure the difference 
\begin{eqnarray}
\ATodd \equiv \frac{1}{2} ( A_T - \ATbar )\,,
\label{eqn:acp}
\end{eqnarray}
in which asymmetries arising from FSI cancel. In addition to being $T$-violating, 
$\ATodd\neq 0$ is manifestly \CP-violating. 
The amount of $\CP$ violation is proportional 
to $\sin\phi\cos\delta$, where $\phi$ and $\delta$ 
are the weak and strong phase differences, respectively, 
between at least two amplitudes contributing 
to the decay~\cite{Datta:2003mj}. This is in 
contrast to a $\CP$ asymmetry 
in $D$ and $\Dbar$ partial widths, $\Acp$, which is proportional to 
$\sin\phi\sin\delta$. If the strong phase difference $\delta$ is
small, $\Acp\approx 0$ but $\ATodd$ could significantly 
differ from zero.

Experimentally, $\ATodd$ has been measured for several $D$ decay 
modes~\cite{FOCUS:2005nwz,BaBar:2011gqb,BaBar:2010xrb,LHCb:2014djq,Belle:2017zvp,Belle:2018pcz,Belle:2022gjv}.
These measurements include two results for charged $D$ mesons:
$\ATodd({D^+\to\KS\Kp\pip\pim})\!=\!(-1.10\pm 1.09)\%$~\cite{FOCUS:2005nwz} and 
$\ATodd({D_s^+\to\KS\Kp\pip\pim})\!=\!(-1.39\pm 0.84)\%$~\cite{BaBar:2011gqb}. 
Here we measure $\ATodd$ for five additional $D_{(s)}^+$ decays: 
${\DpDsp\to\Kp\Km\pip\piz}$, ${\DpDsp\to\Kp\pim\pip\piz}$, and $\DpCF$.
The first $D^+$ decay and the second $D^+_s$ decay are SCS;
the second $D^+$ decay is doubly Cabibbo-suppressed (DCS); and 
the the first $D^+_s$ decay and last $D^+$ decay are 
Cabibbo-favored (CF). 
These are the first such measurements, and we also perform them in seven subregions of phase space.  
In these different subregions, 
the decays proceed predominantly via intermediate processes such as 
${D^+\to\phi\rho^+}$, $\Kstarzb K^{*+}$, 
and $\Kstarzb\rho^+$; and ${D_s^+\to \Kstarp\rho^{0}}$, $\Kstarz\rho^{+}$, 
$\phi\rho^+$, and $\Kstarzb\Kstarp$. As the intermediate states give rise
to different strong phase differences $\delta$~\cite{KANG2010137}, the 
subregions can have different values of~$\ATodd$.

\section{Belle detector and data sets}
\label{sec:data}

This measurement is based on the full data set of the Belle experiment, 
which corresponds to an integrated luminosity of 980~$\invfb$ 
collected at or near the $\Upsilon(nS)$ ($n\!=\!1,\,2,\,3,\,4,\,5$) resonances.
The Belle experiment ran at the KEKB energy-asymmetric collider~\cite{KEKB}.
The Belle detector is a large-solid-angle magnetic spectrometer consisting of a silicon vertex detector~(SVD), 
a $50$-layer central drift chamber~(CDC), 
an array of aerogel threshold Cherenkov counters~(ACC), 
a barrel-like arrangement of time-of-flight scintillation 
counters~(TOF), and an electromagnetic calorimeter~(ECL) 
comprising CsI(Tl) crystals located inside a superconducting 
solenoid coil providing a $1.5$~T magnetic field. 
An iron flux return yoke located outside the coil is instrumented 
to detect $K_L^0$ mesons and to identify muons~(KLM). 
A detailed description of the detector is given in 
Ref.~\cite{belle_detector}. 

Monte Carlo (MC) simulated events are generated from 
$e^+e^-$ collisions based on \textsc{Pythia}~\cite{Sjostrand:2000wi} 
and {\textsc{EvtGen}}~\cite{Lange:2001uf}. All particles are propagated 
through a full detector simulation using {\sc Geant3}~\cite{Brun:1987ma}. 
Final-state radiation from charged particles is simulated using 
{\sc Photos}~\cite{bib:PHOTOS}. 
The MC samples include 
$e^+e^-\!\to q\overline{q}~(q=u,\,d,\,s,\,c)$ and
$e^+e^-\!\to B\overline{B}$ events, and 
$e^+e^-\!\to B_{(s)}^{(*)}\overline{B}_{(s)}^{(*)}X$ events produced at the $\Upsilon(5S)$ resonance. 
These MC samples correspond to an integrated luminosity three times 
that of the data, and they are used to optimize event selection 
criteria, study backgrounds, and evaluate systematic uncertainties.

\section{Event selection}

We reconstruct $D^+$ decays into three final states, 
$\Kp\Km\pip\piz$, $\Km\pip\pip\piz$, and $\Kp\pim\pip\piz$; 
and $D_s^+$ decays into two final states, $\Kp\pim\pip\piz$ and $\Km\Kp\pip\piz$. 
The selection criteria for the Cabibbo-suppressed decays 
are the same as those used to measure their branching fractions
in Ref.~\cite{Belle:2022aha}. The criteria for the CF decays are
separately optimized for the $A_T$ measurement presented here.
Charged tracks are reconstructed 
in the CDC and SVD~\cite{belle_detector}.
To ensure that tracks are well reconstructed, each final-state charged particle 
is required to have at least two SVD hits in each of the longitudinal and azimuthal 
measuring coordinates. We identify $K^\pm$ and $\pi^\pm$ candidates based on particle-identification likelihoods calculated for different
particle hypotheses, $\mathcal{L}_i$ ($i\!=\!\pi,\,K$).
These likelihoods are calculated based 
on information from the ACC, CDC, and TOF~\cite{Nakano:2002jw}.
Tracks with $\mathcal{L}_K/(\mathcal{L}_K\!+\!\mathcal{L}_{\pi})\!>\!0.6$ are 
identified as kaons; otherwise, tracks are considered as pions.
To suppress background from $\DpDsp$ semileptonic decays, tracks that 
are highly electron-like [$\mathcal{L}_e/(\mathcal{L}_{e}\!+\!\mathcal{L}_{\rm hadron})\!>\!0.95$]
or muon-like [$\mathcal{L}_{\mu}/(\mathcal{L}_{\mu}\!+\!\mathcal{L}_{\pi}\!+\!\mathcal{L}_{K})\!>\!0.95$] 
are rejected, 
where these likelihoods
are calculated based on information 
mainly from the ECL and KLM, respectively~\cite{Hanagaki:2001fz,Abashian:2002bd}.
These requirements have overall efficiencies 
of about 90\% for kaons and 95\% for pions.

We reconstruct $\pi^0$ candidates from pairs of photons.
Photon candidates are identified from energy clusters in the ECL 
that are not associated with any charged track. The photon energy
is required to be greater than 50~MeV if reconstructed in the barrel 
region (covering the polar angle $32^\circ\!<\!\theta < 129^\circ$), 
and greater than 100~MeV if reconstructed in the endcap regions
($12^\circ\!<\! \theta\!<\!31^\circ$ or $132^\circ\!<\!\theta\!<\!157^\circ$). 
The energy deposited in the 3$\times$3 array of crystals 
centered on the crystal with the highest energy is 
compared to the energy deposited in the 
corresponding 5$\times$5 array of crystals;
the ratio of the first energy to the second is required to be greater than~0.80. 
Candidate ${\piz\to\gamma\gamma}$ decays are reconstructed 
from photon pairs having an invariant mass satisfying 
$115~{\rm MeV}/c^2 \!<\! M(\gamma\gamma) \!<\! 150~{\rm MeV}/c^2$; 
this region corresponds to about three standard deviations
($\sigma$) in $M(\gamma\gamma)$ resolution. 
The $\piz$ momentum is required to be greater than 0.40~GeV/$c$.
To suppress backgrounds such as $\DpDsp\to\Kp\KS\piz$, we veto 
candidates in which a final-state $\pip\pim$ pair has an 
invariant mass within 10 ${\rm MeV}/c^2$ ($3\sigma$) of 
the $\KS$ mass~\cite{bib:PDG2022}.

A $\DpDsp$ candidate is reconstructed by combining $\Km\Kp\pip$ or $K^{\pm}\pi^{\mp}\pip$ 
track combinations with a $\piz$ candidate. A vertex fit is performed for the three charged 
tracks, and the resulting fit quality is defined as $\chi^2_{\rm vtx}$.
The coordinates of the fitted vertex are assigned as the $\DpDsp$ decay vertex position. 
The momenta of the two photons from the $\piz$ candidate are then fit to this vertex position, 
with the invariant mass constrained to the nominal $\piz$ mass~\cite{bib:PDG2022}; the 
resulting fit quality ($\chi^2_{\piz}$) is required to satisfy $\chi^2_{\piz}<8$.
The $\DpDsp$ production vertex is determined by fitting the 
$\DpDsp$ trajectory to the $e^+e^-$ interaction point (IP), 
which is determined from the beam profiles. This vertex fit 
quality is defined as $\chi^2_{\rm IP}$.

We calculate the significance of the $\DpDsp$ decay length $L/\sigma_L$, 
where $L$ is the projection of the vector running from 
the production vertex to the $\DpDsp$ decay vertex
onto the momentum direction. The corresponding uncertainty 
$\sigma_L$ is calculated by propagating uncertainties in 
the vertices and the $\DpDsp$ momentum, including their correlations.
We subsequently require  
${\chi^2_{\rm vtx} + \chi^2_{\rm IP} < 14}$ and 
${L/\sigma_{\scriptscriptstyle L} > 4}$ for $\DpCS$ decays;
${\chi^2_{\rm vtx} + \chi^2_{\rm IP} < 10}$ and 
${L/\sigma_{\scriptscriptstyle L} > 9}$ for $\DpDCS$ decays;
${\chi^2_{\rm vtx} + \chi^2_{\rm IP} < 10}$ and 
${L/\sigma_{\scriptscriptstyle L} > 2.5}$ for $\DsCS$ decays;
${\chi^2_{\rm vtx} + \chi^2_{\rm IP} < 17}$ and 
${L/\sigma_{\scriptscriptstyle L} > 3}$ for $\DpCF$ decays;
and ${\chi^2_{\rm vtx} + \chi^2_{\rm IP} < 20}$ and 
${L/\sigma_{\scriptscriptstyle L} > 1}$ for $\DsCF$ decays.
These criteria are obtained by maximizing a figure-of-merit, 
$S/\sqrt{S+B}$, where $S$ and $B$ are the 
numbers of signal and background events expected in a 
signal region $-30~{\rm MeV}/c^2\!<\!(M(D)-m_D)\!<\!20~{\rm MeV}/c^2$.
Here, $M(D)$ is the invariant mass of a reconstructed $\Dp$ or $\Dsp$ 
candidate, and $m_D$ is the known $\Dp$ or $\Dsp$ mass~\cite{bib:PDG2022}. This region corresponds to $\pm2.5\sigma$ in resolution. 
To reduce combinatorial backgrounds further, the $\DpDsp$ momentum in 
the $e^+e^-$ center-of-mass frame is required to be greater than 
2.9~GeV/$c$ for $\DsCS$ decays,
and greater than 2.5~GeV/$c$ for the other decay modes.

After applying all selection criteria, about 10\% of events for 
$D^+$ decay modes and 15\% of events for $D_s^+$ decay modes
have multiple signal candidates. For these events, the average 
multiplicity is about 2.2 candidates per event. Such multiple candidates 
arise from the large number of low momentum $\pi^{\pm}$ tracks and $\pi^0$ 
candidates. We select a single $\DpDsp$ candidate per event by choosing the one 
with the smallest value of the sum $\chi^2_{\piz} + \chi_{\rm vtx}^2 + \chi_{\rm IP}^2$.
Based on MC simulation, this criterion selects the correct signal candidate 
68\% of the time. 

We veto several peaking backgrounds arising from $D^{*+}$ 
decays in which the final-state particles are the same as 
those for the signal modes. A detailed description of 
this veto is given in Ref.~\cite{Belle:2022aha}. The veto requirements reject only 
1\%--3\% of signal decays while reducing the $D^{*+}$ backgrounds to a negligible level.

\section{Measurement of $\ATodd$}

We extract the signal yields via an extended unbinned maximum-likelihood fit to the 
$M(D)$ distributions. The probability density function (PDF) used to describe 
backgrounds is taken to be a third-order polynomial for $\DpCS$ and a second-order 
polynomial for the other decay modes. All parameters of these polynomials are free 
to vary in the fit.

The PDF used to describe signal decays 
is taken to be the sum of three asymmetric Gaussian (AG) functions and a Crystal Ball (CB) function~\cite{Gaiser:1985ix}, 
all having a common mean parameter $\mu$
but with different width parameters $\sigma$. 
An additional term is included to describe final-state radiation (FSR).
This term is the sum of a Gaussian function and a CB function.
All signal parameters are fixed to MC values. 
However, the mean parameters of the Gaussian functions 
have a common shift parameter $\delta_{\mu}$, 
and the widths have a common scaling parameter 
$k_{\sigma}$; these two parameters are floated
in the fit to account for differences between MC simulation and data. 
For the large $\DpCF$ sample, $\delta_{\mu}$ and $k_{\sigma}$ are 
floated separately for $D^+$ and $D^-$ decays;
this improves the fit quality. 
For the other decay modes, the difference in resolution between $D^+_{(s)}$ and 
$D^-_{(s)}$ decays is negligible, and we use the same parameters $\delta_{\mu}$ 
and $k_{\sigma}$ for both samples. 
We fit the $M(D)$ distributions of the data to determine $\delta_\mu$ and 
$k_{\sigma}$ along with parameters of the background PDFs; these
are then fixed for subsequent fits. 

We define $C_T$ as the triple product of the momenta of three charged final-state 
particles in the $D_{(s)}^+$ rest frame. For $D^+\to\Km\Kp\pip\piz$ decays,
$C_T = \mbox{\boldmath $\vec{p}$}_{\Km}\cdot(\mbox{\boldmath $\vec{p}$}_{\Kp}\times\mbox{\boldmath $\vec{p}$}_{\pip})$,
and for the charge-conjugate $D^-\to\Kp\Km\pim\piz$ decay,
$\CTbar = \mbox{\boldmath $\vec{p}$}_{\Kp}\cdot(\mbox{\boldmath $\vec{p}$}_{\Km} \times \mbox{\boldmath $\vec{p}$}_{\pim})$. 
For $\DpCF$ decays, the same-sign pions are distinguished by their momenta: 
$C_T = \mbox{\boldmath $\vec{p}$}_{\Km}\cdot (\mbox{\boldmath $\vec{p}$}_{\pip_h}\times\mbox{\boldmath $\vec{p}$}_{\pip_l})$ 
for $D^+$ decays and 
$\CTbar = \mbox{\boldmath $\vec{p}$}_{\Kp}\cdot (\mbox{\boldmath $\vec{p}$}_{\pim_h}\times\mbox{\boldmath $\vec{p}$}_{\pim_l})$ 
for $D^-$ decays, where $\pi^{}_{h}$ and $\pi^{}_{l}$ 
denote the pions with higher and lower momentum.

\begin{figure*}
  \begin{center}%
  \begin{overpic}[width=0.85\textwidth]{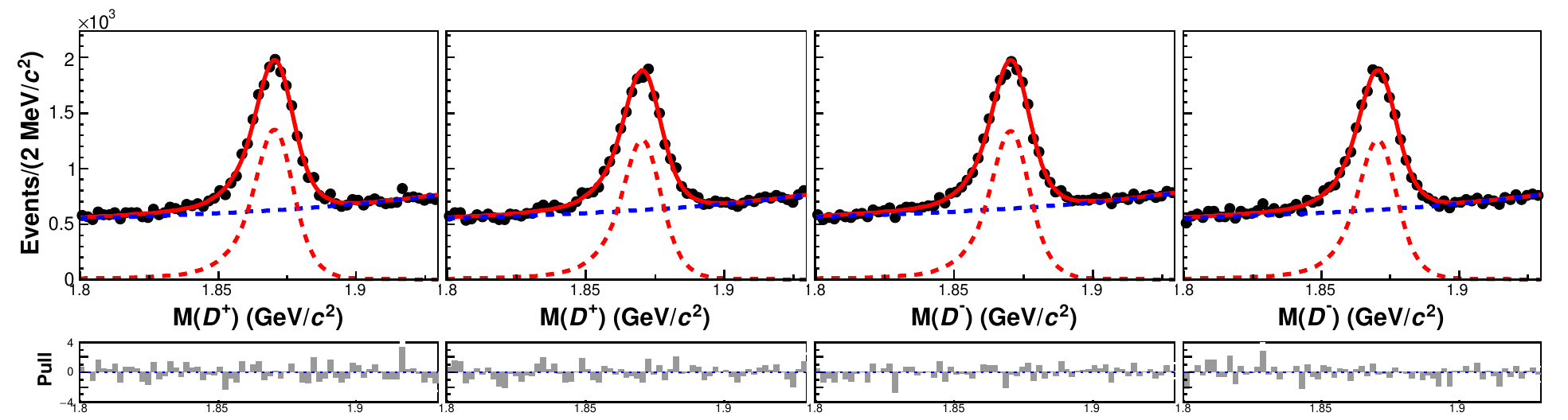}%
  \put(20,26){\small $\DpCS$}%
  \put(67,26){\small $\DpCSb$}%
  \put(7,22){\footnotesize $C_T>0$}%
  \put(30,22){\footnotesize $C_T<0$}%
  \put(54,22){\footnotesize $-\CTbar>0$}%
  \put(77,22){\footnotesize $-\CTbar<0$}%
  \end{overpic}\\
  \vskip13pt
  \begin{overpic}[width=0.85\textwidth]{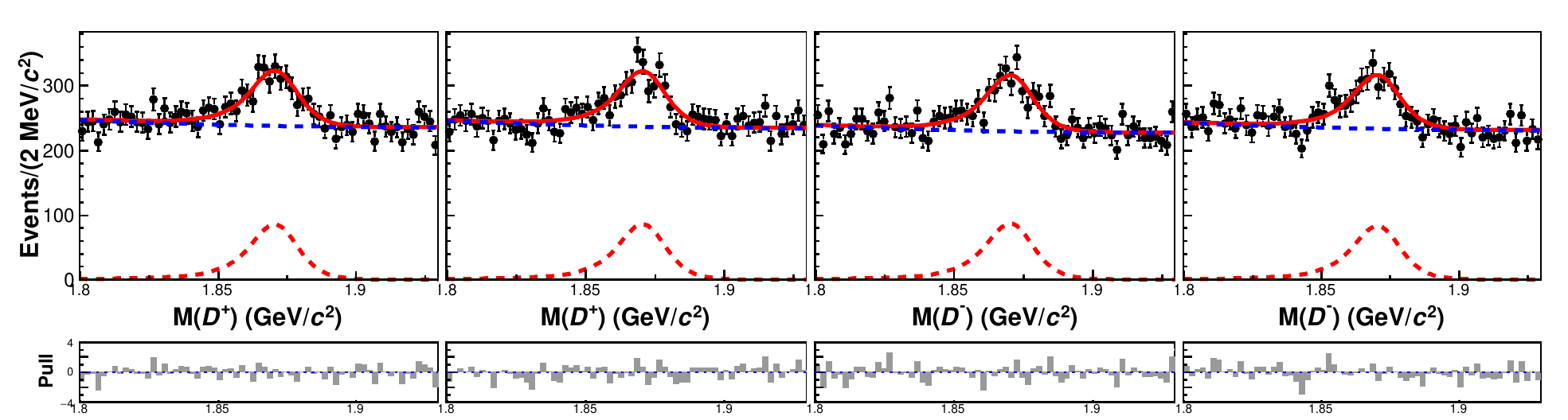}%
  \put(20,26){\small $\DpDCS$}%
  \put(67,26){\small $\DpDCSb$}%
  \end{overpic}\\
  \vskip13pt  
  \begin{overpic}[width=0.85\textwidth]{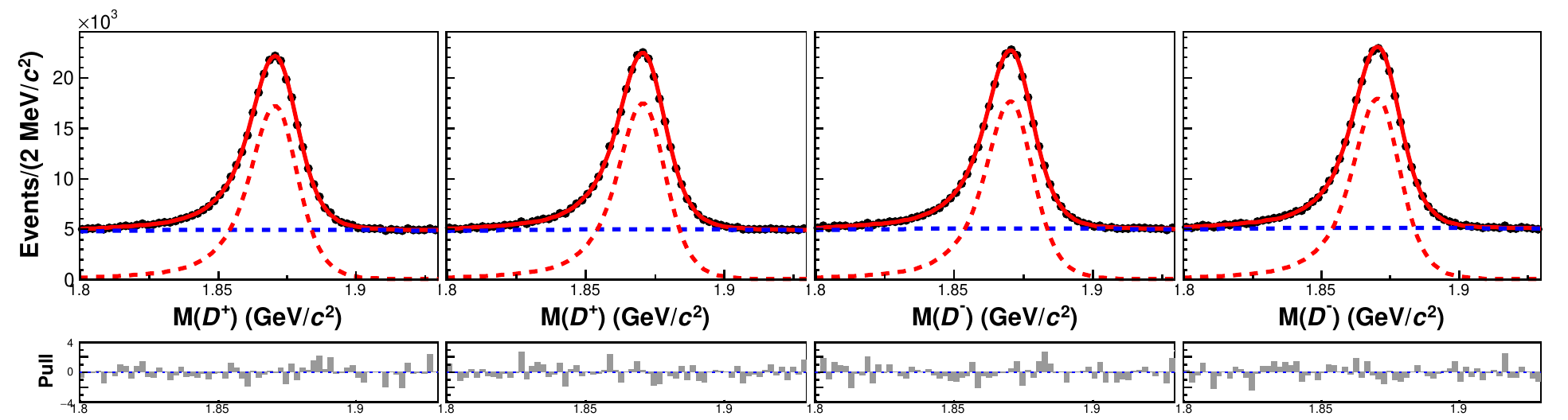}%
  \put(20,26){\small $\DpCF$}%
  \put(67,26){\small $\DpCFb$}%
  \end{overpic}\\
  \vskip13pt  
  \begin{overpic}[width=0.85\textwidth]{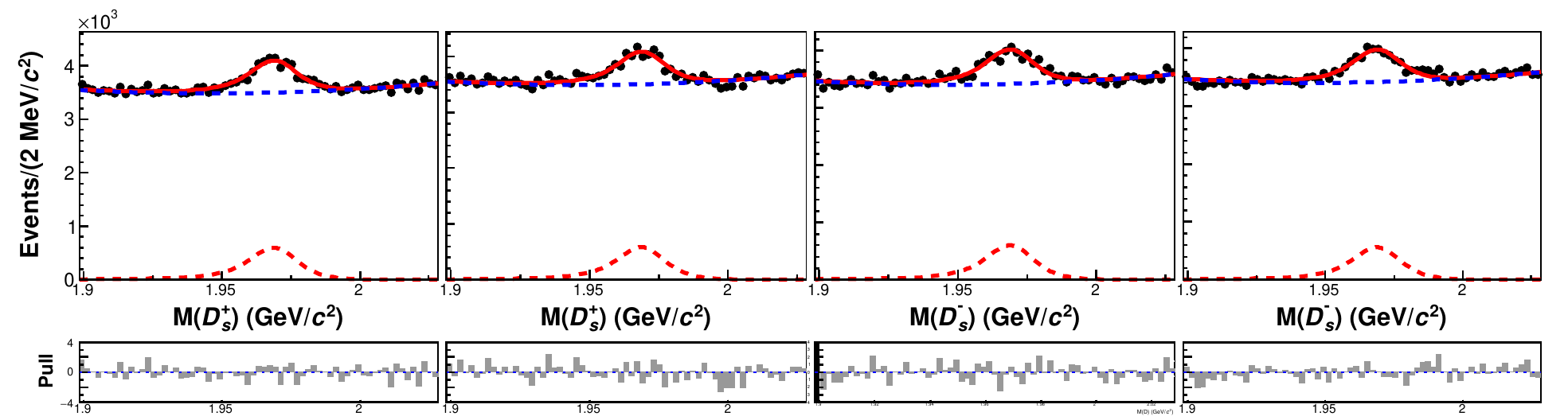}%
  \put(20,26){\small $\DsCS$}%
  \put(67,26){\small $\DsCSb$}%
  \end{overpic}\\
  \vskip13pt  
  \begin{overpic}[width=0.85\textwidth]{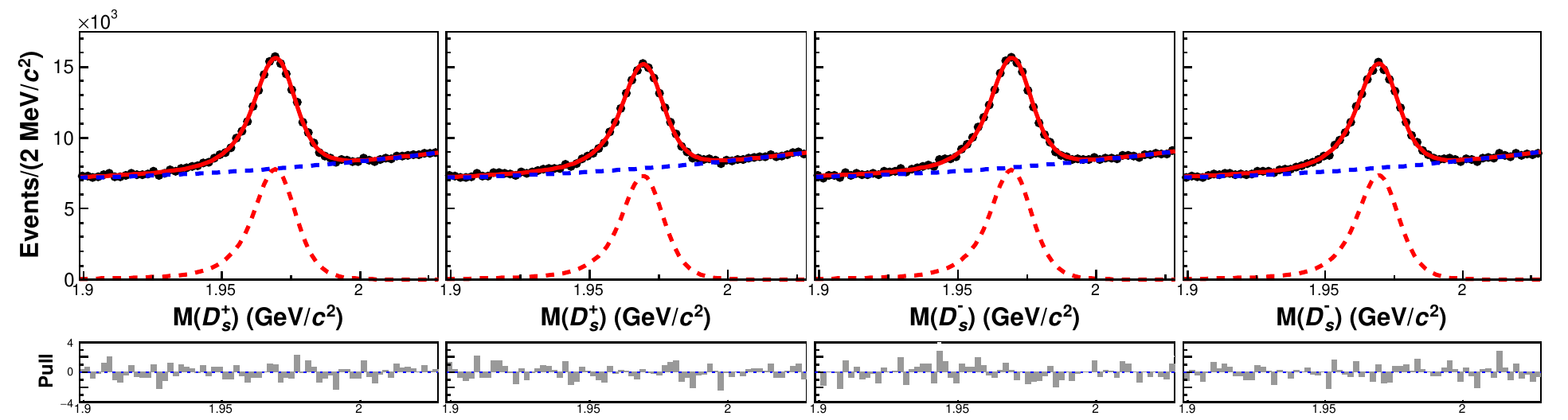}%
  \put(20,26){\footnotesize $\DsCF$}%
  \put(67,26){\footnotesize $\DsCFb$}%
  \end{overpic}\\  
  \caption{\label{fig:ATodd_FitPlots}
  Mass distributions $M(D_{(s)}^{\pm})$ for signal candidates and 
  the fit results. The rows correspond to the decay modes listed. The columns correspond to the four subsamples $D^+_{(s)}$ with $C_T>0$,  $D^+_{(s)}$ with $C_T<0$, $D_{(s)}^-$ with $-\CTbar>0$, and $D^-_{(s)}$ with $-\CTbar<0$, respectively. 
  The points with error bars show the data; the dashed blue curve shows the fitted background;
  the dashed red curve shows the fitted signal; and the solid red curve shows the overall fit result. The corresponding pull distributions are shown below, where the pull is defined as $({\rm data}-{\rm fit})/({\rm uncertainty\ in\ data})$.}
  \end{center}
\end{figure*}
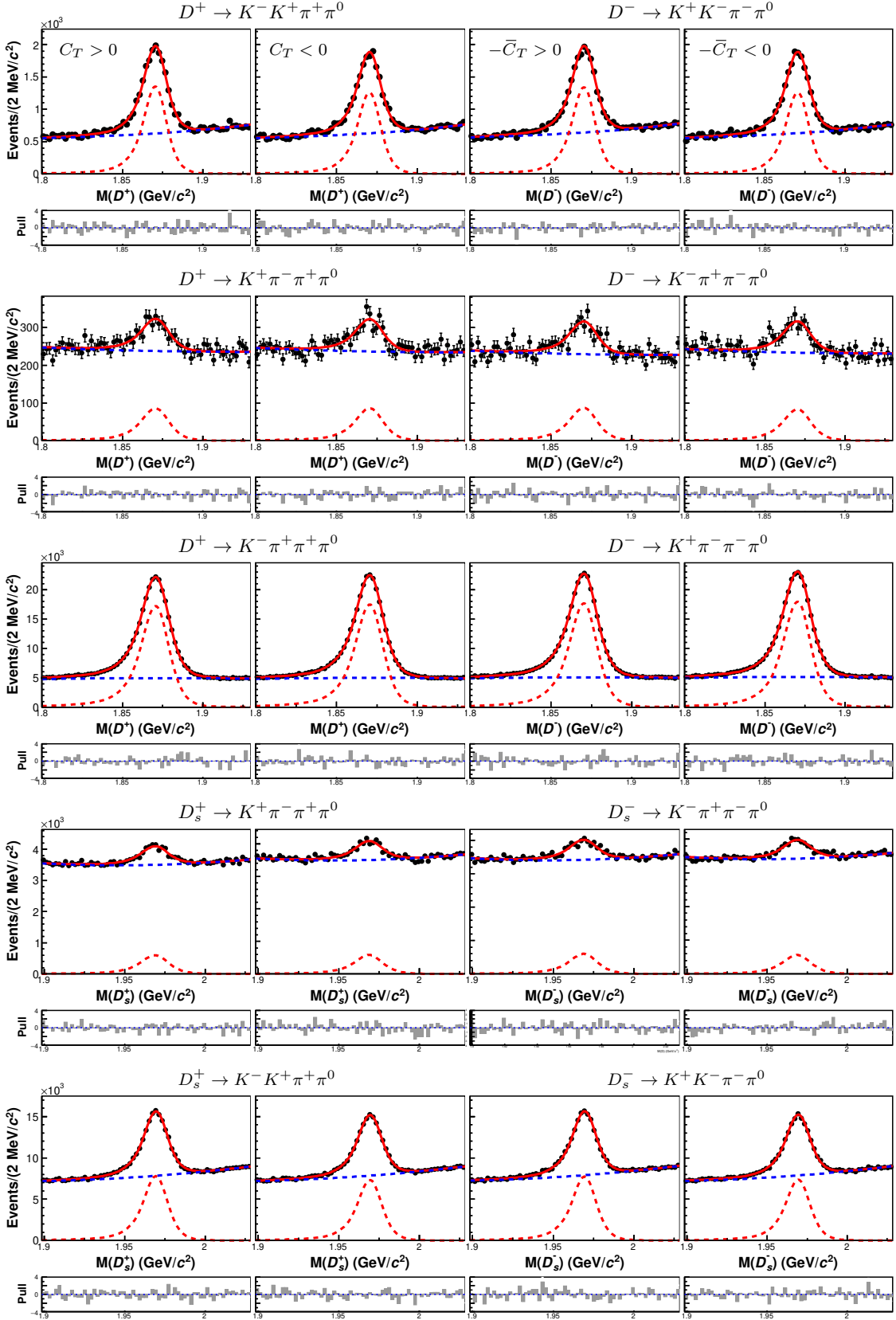

We divide the $D_{(s)}^{\pm}$ candidates into four subsamples 
based on the $D_{(s)}^{\pm}$ charge and the sign of $C_T$ or $\CTbar$:
(1) $D_{(s)}^+$ with $C_T\!>\!0$, 
(2) $D_{(s)}^+$ with $C_T\!<\!0$, 
(3) $D_{(s)}^-$ with $-\overline{C}_T\!>\!0$, and 
(4) $D_{(s)}^-$ with $-\overline{C}_T\!<\!0$.
Subsamples (1) and (3) are related by a $\CP$ transformation, as are subsamples (2) and~(4).
The asymmetries $A_T$ and $\overline{A}_T$ are defined in terms of these signal yields
in Eq.~(\ref{eqn:AT}), and the $T$-violating asymmetry $\ATodd$ is defined in
terms of $A_T$ and $\overline{A}_T$ in Eq.~(\ref{eqn:acp}). We fit these four 
subsamples simultaneously, taking the independent fitted parameters to be 
$N^{}_{D} = N(C_T\!>\!0) + N(C_T\!<\!0)$,
$N^{}_{\Db} = \overline{N}(-\CTbar\!>\!0) + \overline{N}(-\CTbar\!<\!0)$,
$A_T$, and $\ATodd$.
We test this fitting procedure on 
MC samples and obtain fitted values for $\ATodd$ 
always consistent with the input values within 
their statistical uncertainties.
The results of the fits to data
are plotted in Fig.~\ref{fig:ATodd_FitPlots} and
listed in Table~\ref{tab:ToddFitFinal}.
The plots show good agreement between the data and the fitted functions.

\begin{table*}
\begin{center}
\caption{\label{tab:ToddFitFinal}
Fitted $D^+_{(s)}$ and $D^-_{(s)}$ signal yields ($N_{D}$ and $N_{\Db}$, respectively), 
$T$-odd asymmetry $A_T$, and $T$-odd $\CP$-violating asymmetry $\ATodd$, obtained from a
simultaneous fit to the four subsamples of each decay mode. 
The uncertainties listed are statistical only.}
\renewcommand\arraystretch{1.2}
\setlength{\tabcolsep}{4mm}{
\begin{tabular}{c|ccc|cc} \hline \hline   
Decay   & \multicolumn{3}{c|}{$D^+\to f$} & \multicolumn{2}{c}{$D^+_s\to f$} \\
\hline
 Final state ($f$)  & $K^+K^-\pi^+\pi^0$  &  $K^+\pi^-\pi^+\pi^0$ & $K^-\pi^+\pi^+\pi^0$  &  
$K^+\pi^-\pi^+\pi^0$  &  $K^+K^-\pi^+\pi^0$ \\ 
\hline    
$N_{D}$ & $27284\pm 254$  & $2062\pm 127$	& $438432\pm 947$ & $15197\pm 484$ & $167357\pm 786$	\\
$N_{\Db}$ & $27177\pm 255$	& $2044\pm 125$	& $450667\pm 961$ & $14945\pm 479$ & $167064\pm 788$	\\
$A_T$ (\%) & $+3.63\pm 0.93$	& $-0.4\pm 6.0$	& $-0.76\pm 0.22$ & $+1.4\pm 3.2$    & $+2.96\pm 0.47$	\\
$\ATodd$ (\%) & $+0.26\pm 0.66$	& $-1.3\pm 4.2$	& $+0.02\pm 0.15$  & $-1.1\pm 2.2$  & $+0.22\pm 0.33$ \\ 
\hline  \hline 
\end{tabular}
}
\end{center}  
\end{table*}

\section{Measurements of $\ATodd$ in subregions of phase space}

\begin{figure*}
  \begin{center}%
  \begin{overpic}[width=0.85\textwidth]{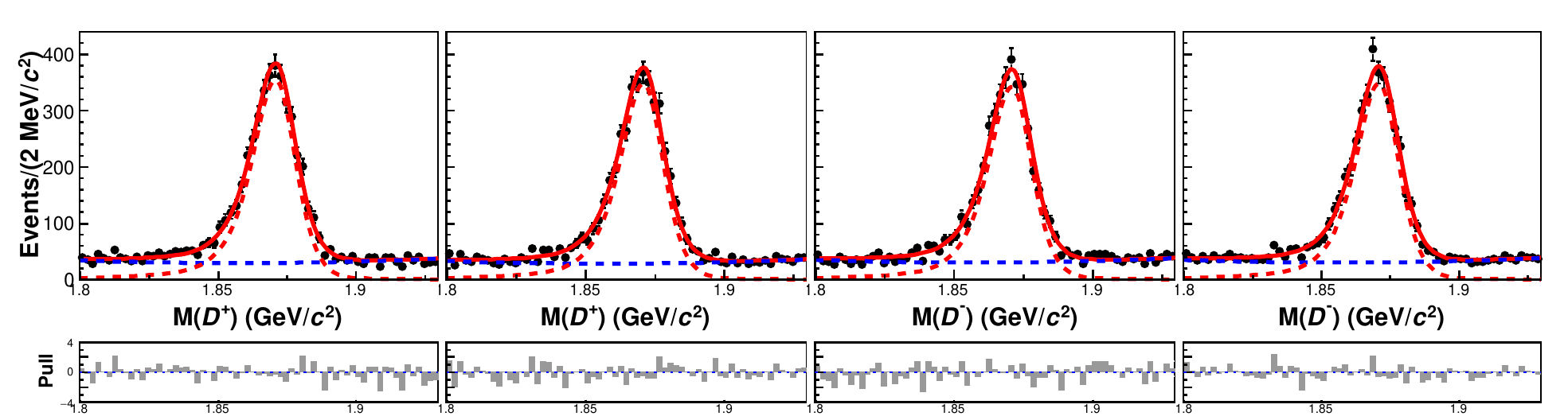}%
  \put(24,26){\small $D^+\to\phi\rho^+$}%
  \put(71,26){\small $D^-\to\phi\rho^-$}%
  \put(7,22){\footnotesize $C_T>0$}%
  \put(30,22){\footnotesize $C_T<0$}%
  \put(54,22){\footnotesize $-\CTbar>0$}%
  \put(77,22){\footnotesize $-\CTbar<0$}%
  \end{overpic}\\
  \vskip13pt
  \begin{overpic}[width=0.85\textwidth]{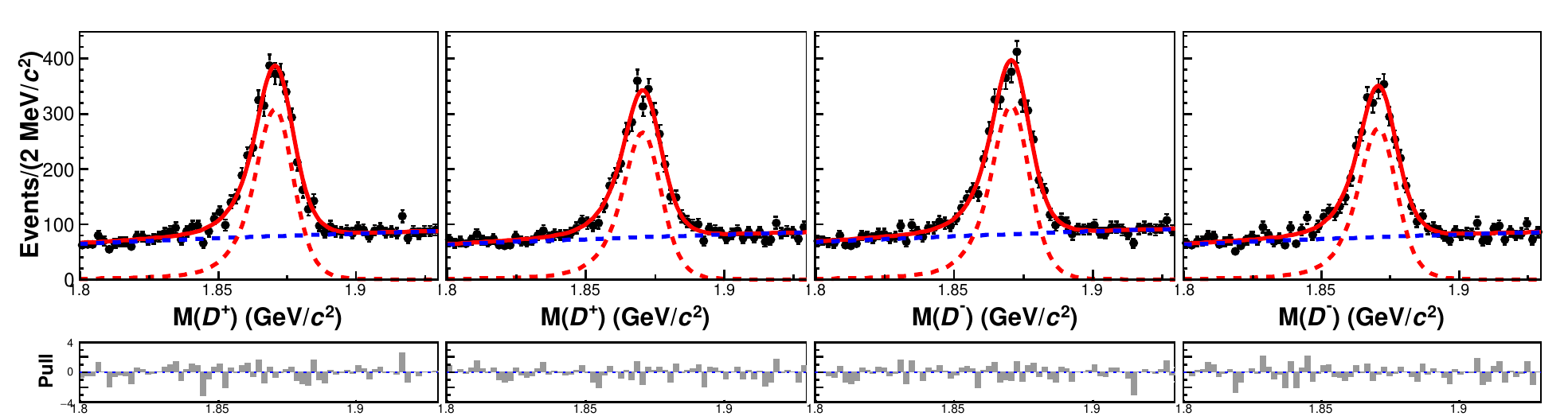}%
  \put(24,26){\small $D^+\to\bar{K}^{*0} K^{*+}$}%
  \put(71,26){\small $D^-\to K^{*0} K^{*-}$}%
  \end{overpic}\\
  \vskip13pt
  \begin{overpic}[width=0.85\textwidth]{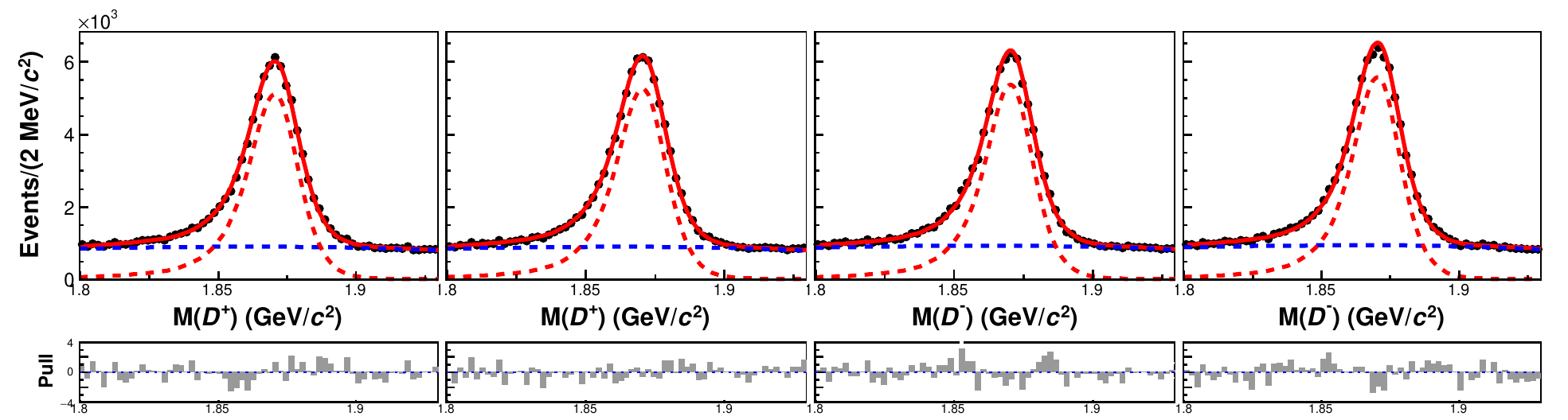}%
  \put(24,26){\small $D^+\to \bar{K}^{*0}\rho^+$}%
  \put(71,26){\small $D^-\to K^{*0}\rho^-$}%
  \end{overpic}  
  \caption{\label{fig:Todd4subDpFinal}
  Mass distributions $M(D^{\pm})$ and fit results for subregions of phase space. 
  The rows correspond to the decay modes listed. The columns correspond 
  to the four subsamples
  $D^+$ with $C_T>0$,  $D^+$ with $C_T<0$, $D^-$ with $-\CTbar>0$, and $D^-$ with $-\CTbar<0$, 
  respectively. The points with error bars show the data; the dashed blue curve shows the fitted background;
  the dashed red curve shows the fitted signal; and the solid red curve shows the overall fit result. The corresponding pull distributions are shown below.
  }
  \end{center}
\end{figure*}

\begin{figure*}
  \begin{center}%
  \begin{overpic}[width=0.85\textwidth]{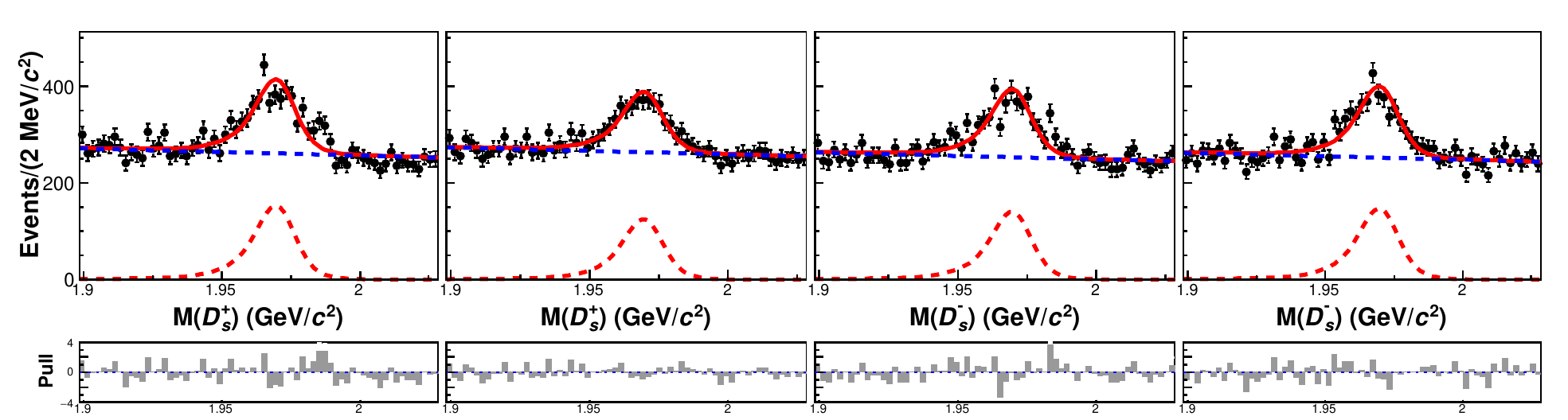}%
  \put(24,26){\small $D_s^+\to \bar{K}^{*0}\rho^{+}$}%
  \put(71,26){\small $D_s^-\to K^{*0}\rho^{-}$}%
  \put(7,22){\footnotesize $C_T>0$}%
  \put(30,22){\footnotesize $C_T<0$}%
  \put(54,22){\footnotesize $-\CTbar>0$}%
  \put(77,22){\footnotesize $-\CTbar<0$}%
  \end{overpic}\\
  \vskip13pt
  \begin{overpic}[width=0.85\textwidth]{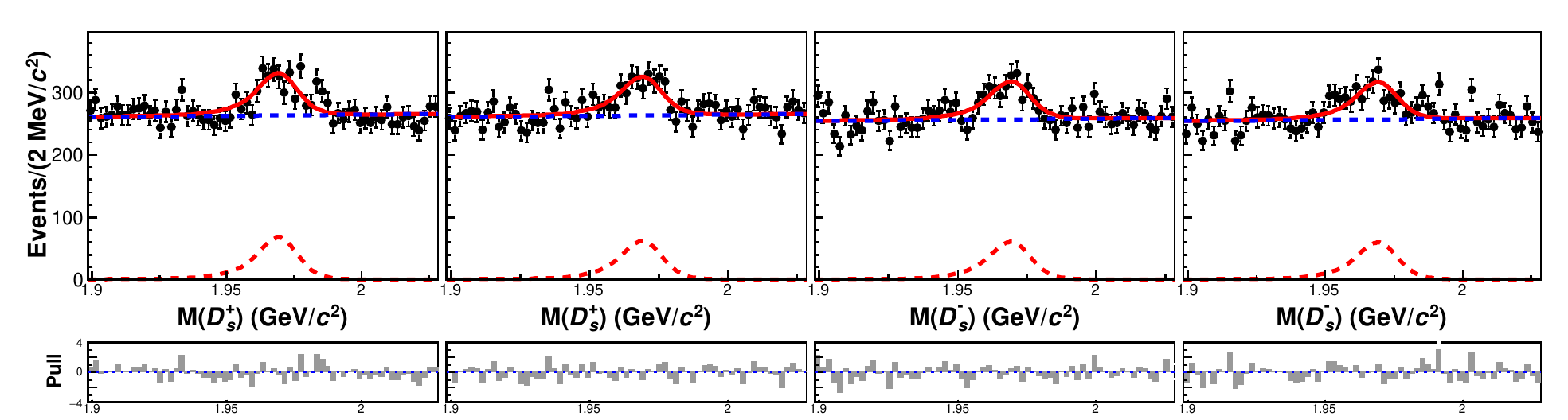}%
  \put(24,26){\small $D_s^+\to K^{*+}\rho^{0}$}%
  \put(71,26){\small $D_s^-\to K^{*-}\rho^{0}$}%
  \end{overpic}\\
  \vskip13pt
  \begin{overpic}[width=0.85\textwidth]{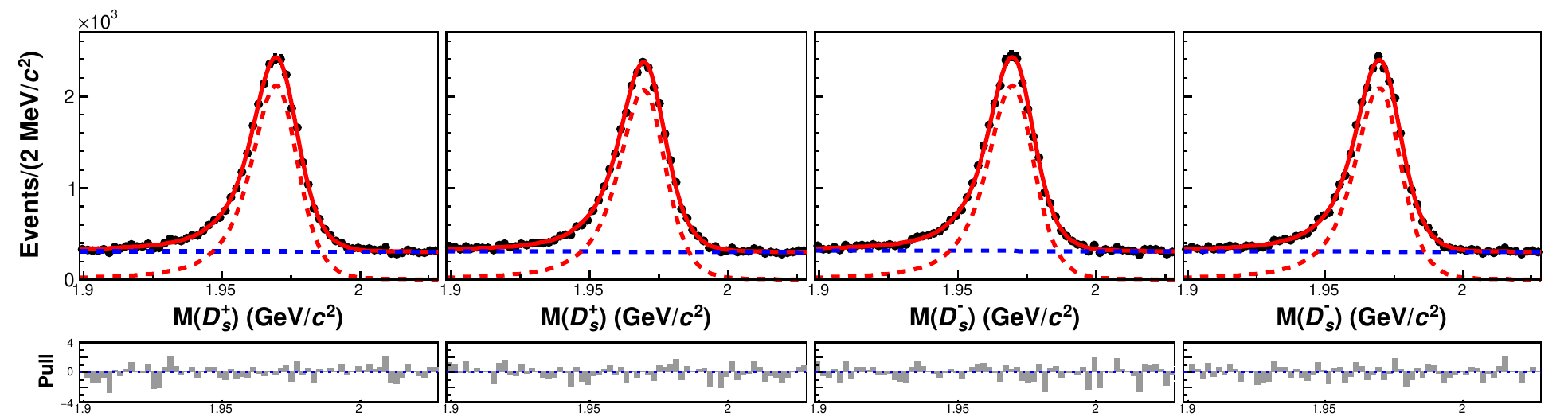}%
  \put(24,26){\small $D_s^+\to\phi\rho^+$}%
  \put(71,26){\small $D_s^-\to\phi\rho^-$}%
  \end{overpic}\\
  \vskip13pt
  \begin{overpic}[width=0.85\textwidth]{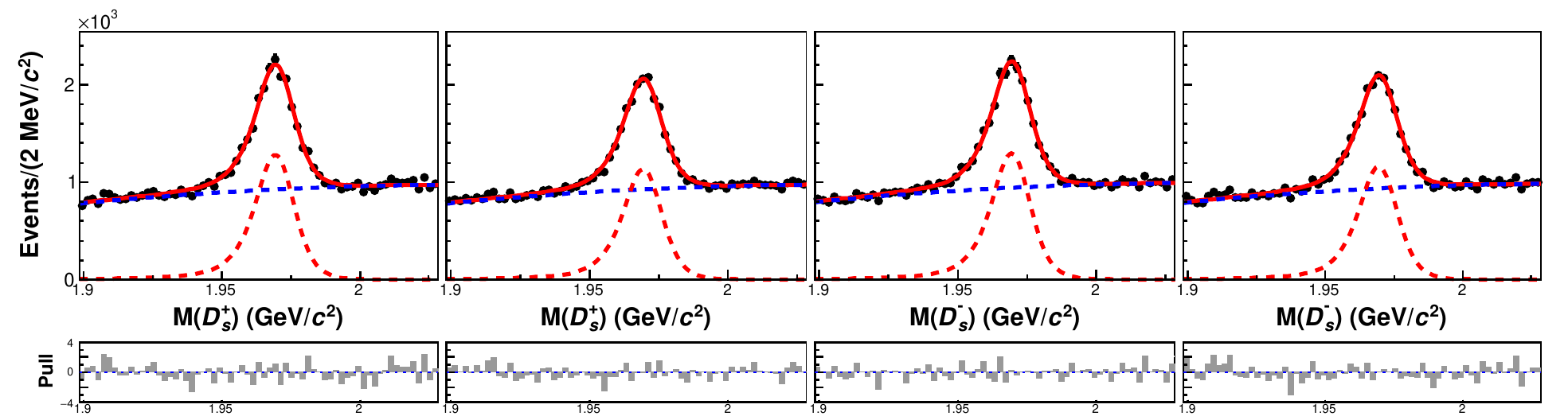}%
  \put(24,26){\small $D_s^+\to\bar{K}^{*0}K^{*+}$}%
  \put(71,26){\small $D_s^-\to K^{*0}K^{*-}$}%
  \end{overpic}
  \caption{\label{fig:Todd4subDsFinal}
  Mass distributions $M(D_{s}^{\pm})$ and 
  fit results for subregions of phase space. 
  The rows correspond to the decay modes listed. The columns correspond to the four subsamples
  $D^+_{s}$ with $C_T>0$,  $D^+_{s}$ with $C_T<0$, $D_{s}^-$ with $-\CTbar>0$, and $D^-_{s}$ with $-\CTbar<0$, 
  respectively. The points with error bars show the data; the dashed blue curve shows the fitted background;
  the dashed red curve shows the fitted signal; and the solid red curve shows the overall fit result. The corresponding pull distributions are shown below.
  }
  \end{center}
\end{figure*}

The distributions of invariant mass for pairs of 
final-state particles exhibit significant peaks
corresponding to intermediate states 
$\phi(1020)$, $K^{*}(892)^{0,+}$, and $\rho(770)$. These 
states give rise to different strong phases among decay
amplitudes, and this can lead to different amounts of direct $\CP$
violation. Thus, we search for $\CP$ violation in ranges of 
invariant mass corresponding to a decay proceeding through 
intermediate resonances $\phi$, $\rho$, or $K^*$:
$|M_{KK}-m_{\phi}|<10~{\rm MeV}/c^2$, 
$-90~{\rm MeV}/c^2<(M_{\pip\piz}-m_{\rho})<60~{\rm MeV}/c^2$, 
and $|M_{K\pi} - m_{K^{*}}|<60~{\rm MeV}/c^2$.
The requirements for seven different
intermediate decays are listed in Table~\ref{tab:ATodd4DToVV}.
We analyze these regions only for SCS and CF decays, as
the signal yield for the DCS decay $\DpDCS$ is too low. 
The subregions of invariant mass correspond to the 
$D^{+}\to K^{\pm} h^{\mp}\pi^{+}\pi^{0}$ decay proceeding via 
$D^+\to\phi\rho^+$, $\Kstarzb K^{*+}$, and $\Kstarzb\rho^+$; 
and the $D^{+}_s\to K^{+} h^{-}\pi^{+}\pi^{0}$ decay proceeding via 
$D_s^+\to K^{*+}\rho^{0}$, $K^{*0}\rho^{+}$, $\phi\rho^+$, and $\Kstarzb K^{*+}$.
For the $D_{(s)}^+\to\Kstarzb\Kstarp$ subsamples, 
we additionally require $|M_{KK}-m_{\phi}|>10~{\rm MeV}/c^2$
in order to remove contamination from 
$D_{(s)}^+\to\phi\pi^+\pi^0$ decays.
For this measurement, we use the same fitting procedure  
as used for the $\ATodd$ measurement over the entire phase space. 
For each invariant-mass subregion, we first perform an $M(D)$ fit to obtain 
the shape of signal and background. 
The floated parameters of the signal PDF ($\delta^{}_\mu$ and $k^{}_\sigma$) and the background PDF are allowed to 
differ among different subregions.
We subsequently perform a simultaneous fit to the four 
$C_T$ samples. The resulting values of
$\ATodd$ are listed in Table~\ref{tab:ATodd4DToVV}, and projections of the fits are shown in 
Fig.~\ref{fig:Todd4subDpFinal} for $D^+$ decays and in Fig.~\ref{fig:Todd4subDsFinal} for 
$D_s^+$ decays. 
The subregion dominated by $D_s^+\to K^{*0}\rho^{+}$ has the largest $\CP$ asymmetry:
$\ATodd=(6.2\pm 3.0\pm 0.4)\%$,
which differs from zero by $2\sigma$.

\begin{table*}[!htbp]
\begin{centering}
\caption{\label{tab:ATodd4DToVV}$T$-odd $\CP$-violating 
asymmetries ($\ATodd$) in seven subregions of phase space (see text)
corresponding to the intermediate $D_{(s)}^+\to VV$ processes listed.
The uncertainties listed are statistical and systematic, respectively.}
\setlength{\tabcolsep}{4mm}{
\renewcommand\arraystretch{1.2}
\begin{tabular}{llll} \hline \hline
Subregion & $D^+_{(s)}\to VV$ 		& Signal region (SR)  		&   $\ATodd$ ($\times10^{-2}$)    \\ 
\hline   
(1) SCS	& $\Dp\to \phi\rho^+$   	&  $\phi$-SR, $\rho^+$-SR    	&   $0.85\pm 0.95\pm 0.25$  \\
(2) SCS	& $\Dp\to \Kstarzb K^{*+}$    & $K^{*(0,+)}$-SR, veto $\phi$-SR  &   $0.17\pm 1.26\pm 0.13$ \\
(3) CF	& $\Dp\to \Kstarzb\rho^+$    & $K^{*0}$-SR,  $\rho^+$-SR  	&   $0.25\pm 0.25\pm 0.13$   \\ 
(4) SCS	& $D_s^+\to K^{*0}\rho^{+}$    &  $K^{*0}$-SR,  $\rho^+$-SR   	&   $6.2\,\,\,\pm 3.0\,\,\,\pm 0.4\,\,\,$   \\
(5) SCS	& $D_s^+\to K^{*+}\rho^{0}$    &  $K^{*+}$-SR,  $\rho^0$-SR   	&   $1.7\,\,\,\pm 6.1\,\,\,\pm 1.5\,\,\,$   \\
(6) CF	& $D_s^+\to \phi\rho^+$    & $\phi$-SR, $\rho^+$-SR 		&   $0.31\pm 0.40\pm 0.43$   \\
(7) CF	& $D_s^+\to \Kstarzb K^{*+}$  &  $K^{*(0,+)}$-SR,  veto $\phi$-SR &   $0.26\pm 0.76\pm 0.37$   \\  
\hline \hline 
\end{tabular}}
\end{centering}
\end{table*}

\section{Systematic uncertainties}

Many of the systematic uncertainties in the 
measurements of $C_T$ and $\CTbar$
cancel in the ratios $A_T$ and $\overline{A}_T$ (Eq.~\ref{eqn:AT}) 
or in the difference between them 
(Eq.~\ref{eqn:acp}). The uncertainties that do not cancel are
discussed below and listed in Table~\ref{tab:AcpSys}. For a given decay mode, the
total systematic uncertainty is the sum in quadrature of the individual contributions. 
For our measurements corresponding to subregions of phase space as listed in 
Table~\ref{tab:ATodd4DToVV}, we evaluate systematic uncertainties in an 
identical manner; these results are listed in Table~\ref{tab:AcpSysD2VV}.

\begin{table*}
\begin{centering}
\caption{\label{tab:AcpSys}
Systematic uncertainties for $\ATodd$ in \% for 
five $D_{(s)}^+$ decay channels: 
(a)~$\DpCS$; (b)~$\DpDCS$; (c)~$\DpCF$; (d)~$\DsCS$; 
and (e)~$\DsCF$.}
\setlength{\tabcolsep}{4mm}{
\renewcommand\arraystretch{1.2}
\begin{tabular}{lccccc} 
\hline \hline
Decay channel & (a) & (b) & (c) & (d) & (e) \\
\hline    
$C_T$-dependent efficiency  	&  $0.13$    	& $0.02$   	&  $0.08$  	&  $0.02$  	&  $0.41$ 	\\
$C_T$ resolution 		&  $0.01$ 	&    $0.06$    &   $0.01$  	&    $0.07$ 	&   $0.02$  	\\
PDF parameters &  $0.01$	&  $0.07$	&  $0.01$	& $0.07$  	&  $0.04$	\\
Mass resolution		&  $0.03$  	&  $0.01$	&  ...  	&  $0.02$    	&  $0.11$	\\
Fit bias 			&  $0.01$	&  $0.07$	&  $0.00$	&     $0.06$	&     $0.02$	\\ 
\hline 
Total syst.	&  $0.13$  	&   $0.12$	&    $0.08$	
							&   $0.12$    &  $0.43$		\\ 
\hline \hline 
\end{tabular}}
\end{centering}  
\end{table*}

\begin{table*}
\begin{centering}
\caption{\label{tab:AcpSysD2VV}
Systematic uncertainties for $\ATodd$ in \% for seven 
subregions of phase space as defined in Table~\ref{tab:ATodd4DToVV}. These subregions correspond to various two-body intermediate 
processes.}
\setlength{\tabcolsep}{4mm}{
\renewcommand\arraystretch{1.2}
\begin{tabular}{lccccccc} 
\hline \hline   
Source 	&  (1) 		&  (2)		&   (3)		&    (4) &     (5)	& (6) 	&  (7)		\\ 
\hline 	
$C_T$-dependent efficiency &  $0.24$	&  $0.13$	&  $0.13$	&   $0.29$	&   $1.17$	& $0.43$  & $0.37$ \\
$C_T$ resolution  &   $0.07$ 	&   $0.00$ 	&  $0.00$      &   $0.06$     &   $0.16$    	& $0.05$  & $0.02$ \\
PDF parameters & $0.01$ & $0.02$	& $0.00$	&  $0.09$	&  $0.55$	& $0.00$ & $0.01$    \\
Mass resolution	 &  $0.01$	&   $0.02$	& ...	&  $0.26$	&   $0.80$	& $0.01$  & $0.01$ \\
Fit bias 	 &  $0.03$	&   $0.01$	&  $0.01$	&   $0.03$	&   $0.09$	& $0.00$  & $0.03$ \\ 
\hline 
Total syst.	&   $0.25$  & $0.13$ & $0.13$  & $0.41$	
        &  $1.53$	& $0.43$ & $0.37$	\\ 
\hline \hline 
\end{tabular}}
\end{centering}  
\end{table*}

\subsection*{(1) $C_T$-dependent efficiency}

We study the dependence of the reconstruction efficiency on 
$C_T$ using a large sample of MC signal events. We find 
that events with smaller $|C_T|$ values have lower efficiency due to the fact that the final-state $\piz$ tends to have lower momentum. 
To assign a systematic uncertainty for this, we fit 
the $C_T$-dependent efficiencies with a polynomial function
$\varepsilon(C_T) = \varepsilon_0 + c_{1}\cdot|C_T| + c_{2} \cdot |C_T|^2$. 
This fit is performed separately for the 
$D^+_{(s)}$ subsamples having $C_T\!>\!0$ and $C_T\!<\!0$,
and similarly for $D^-_{(s)}$. 
With these functions, we weight $D_{(s)}^{\pm}$ candidates 
in data by the reciprocals of their efficiencies (normalized 
by the average efficiency) and repeat the fit for $\ATodd$.
The difference between the fit result and the nominal result 
for $\ATodd$, while consistent with zero as expected, is 
conservatively assigned as a systematic uncertainty due to 
a possible $C_T$-dependence of the reconstruction efficiency.

\subsection*{(2) $C_T$-resolution effects}

The resolution in $C_T$ can affect $\ATodd$ if the smearing of events from positive to negative $C_T$ values
differs from the smearing of events from negative to positive values. This is studied using a large 
MC sample to calculate the resolution function for different slices of $C_T$. A double-Gaussian function
is used to model the resolution. The effective width of the double Gaussian is plotted
as a function of $C_T$, and the resulting distribution is fitted to a polynomial $f(x)=a_0+a_1 x+a_2 x^2$.
The fitted value of $a_1$, the coefficient of the asymmetric term, is consistent with zero, and
thus the effects of $C_T$ resolution are expected to be small.

To determine the effect on $\ATodd$, we smear the $C_T$ distributions by Gaussian functions
having widths given by the previously fitted polynomial $a_0+a_1 x+a_2 x^2$. We repeat the 
fit for $\ATodd$ on these smeared samples and take the difference between the resulting value 
of $\ATodd$ and the nominal value as a systematic uncertainty due to $C_T$ resolution.

\subsection*{(3) Signal and background parameters}

We consider systematic uncertainty arising from the fixed parameters 
of signal and background PDFs as follows. 
We sample the parameters of the signal PDF from a multi-dimensional Gaussian 
function, accounting for their respective uncertainties and correlations,
and repeat the fit for $\ATodd$. 
We repeat this procedure 1000 times and take the RMS 
of the fitted $\ATodd$ values as systematic uncertainties:
$0.002\%$ for $\DpCS$, $0.023\%$ for $\DpDCS$, $0.004\%$ for $\DpCF$, 
$0.054\%$ for $\DsCS$, and $0.003\%$ for $\DsCF$.
We calculate the uncertainty due to fixed values of $\delta_\mu$, $k_\sigma$,
and background parameters
in an identical manner;
the resulting uncertainties are $0.010\%$, $0.071\%$, $0.001\%$, 
$0.048\%$, and $0.002\%$, respectively. 
We sum the two uncertainties in quadrature for each 
channel to obtain the overall systematic uncertainty due to 
fixed signal and background parameters.

\subsection*{(4) Signal mass resolution}

We consider a possible difference between $D_{(s)}^+$ and $D_{(s)}^-$ 
samples arising from a difference in mass resolution between data and MC events. 
To study this effect, we fit the $M(D_{(s)}^+)$ and $M(D_{(s)}^-)$ distributions 
individually to obtain separate sets of parameters: $(\delta_{\mu}^+,\,k_{\sigma}^+)$ 
for $D_{(s)}^+$ decays, and $(\delta_{\mu}^-,\,k_{\sigma}^-)$ for $D_{(s)}^-$ decays.
With these parameters, we repeat the simultaneous fit to the $C_T$
subsamples. The resulting change in $\ATodd$ is assigned as a systematic
uncertainty due to signal mass resolution.

\subsection*{(5) Fit bias}

To check for bias in our fitting procedure, we study large samples 
of ``toy'' MC events. These events are generated by 
sampling from the PDFs used to fit the data. We generate 1000 samples 
each for different input values of $\ATodd$, spanning a range from 
$-0.05$ to $+0.05$.
We fit these samples and plot the resulting $\ATodd$ values. This distribution is then fitted with a Gaussian function, and the mean and width of the Gaussian 
are taken as the mean measured value and its uncertainty 
corresponding to that input value of $\ATodd$. Plotting these 
mean values versus the input value displays a linear dependence;
fitting a line to these points gives a slope consistent with unity and an 
intercept consistent with zero. 
However, we conservatively assign the difference between our measured value and the 
input value corresponding to our measured value 
as given by this fitted line (including its uncertainty) 
as a systematic uncertainty due to possible fit bias.

\section{Summary}

\begin{figure}[!hbtp]
  \begin{centering}%
  \begin{overpic}[width=0.52\textwidth]{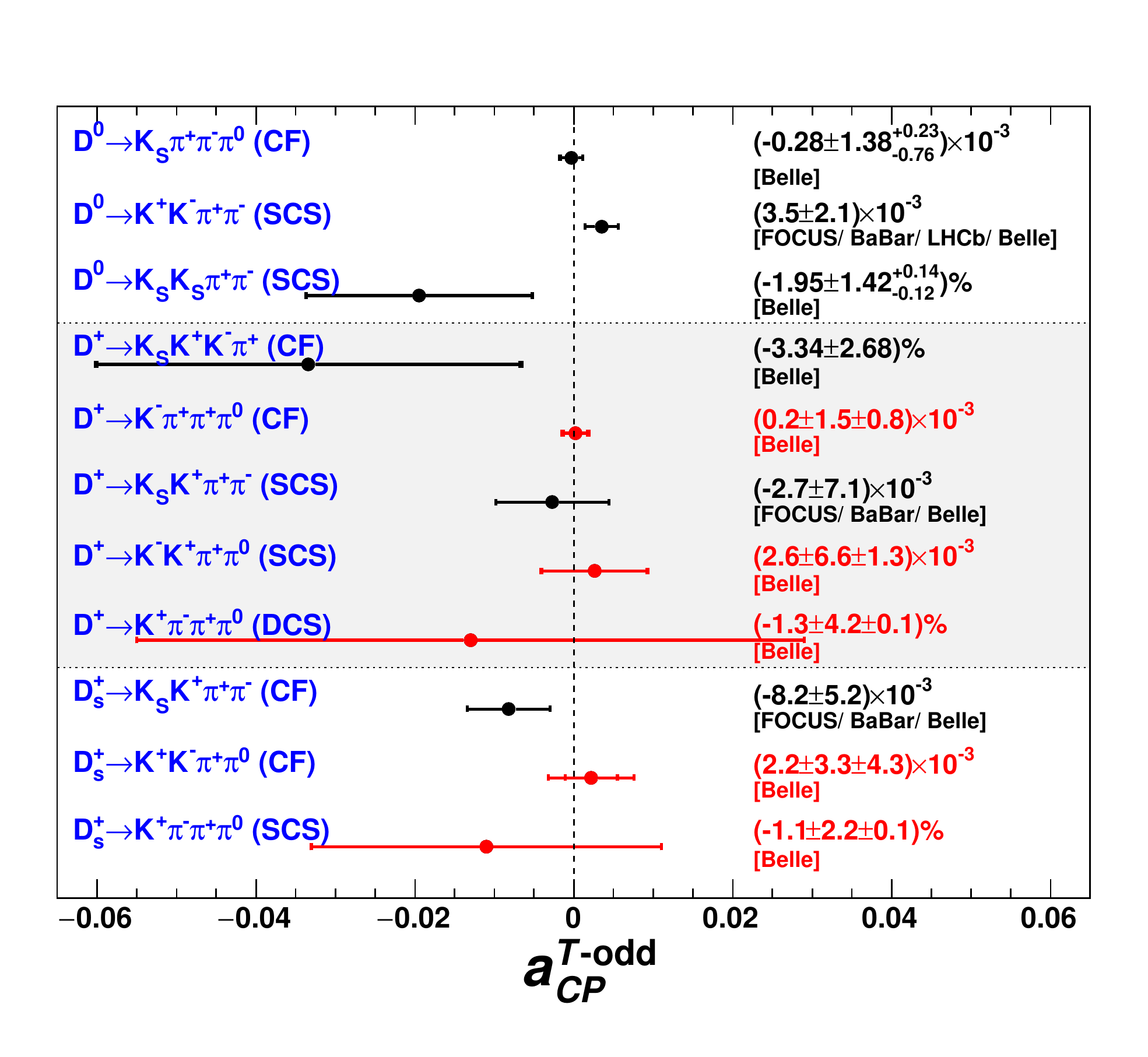}%
  \end{overpic}
  \vskip-20pt
  \caption{\label{fig:resultTodd}
  Our results for $\ATodd$ (in red) along with other measurements of $\ATodd$ for $D^0$ and $D^+_{(s)}$ decays~\cite{HFLAV:2019otj,Belle:2022gjv,Belle:Moon}. For decays in which more than one measurement has been made, the world average value is plotted.}
  \end{centering}
\end{figure}

We perform a search for $\CP$ violation using a $T$-odd observable
for five $D_{(s)}^+$ decays.
Our analysis is based on the full data set of 980~$\invfb$ collected 
by the Belle experiment. Our results for the \CP-violating 
$T$-odd parameter $\ATodd$ are
\begin{eqnarray*}
\ATodd(\DpCS) & = & (+2.6\pm 6.6\pm1.3 )\times10^{-3}  \\
\ATodd(\DpDCS) & = & (-1.3\pm 4.2\pm 0.1 )\times10^{-2}  \\
\ATodd(\DpCF) & = & (+0.2\pm 1.5\pm0.8 )\times10^{-3} \\
\ATodd(\DsCS) & = & (-1.1\pm 2.2\pm 0.1 )\times10^{-2}  \\
\ATodd(\DsCF) & = & (+2.2\pm 3.3\pm4.3 )\times10^{-3}\,, 
\end{eqnarray*} 
where the first uncertainties are statistical and the second uncertainties are systematic.
These values are all consistent with zero and show no evidence of $\CP$ violation.  We plot 
these values in Fig.~\ref{fig:resultTodd} along with other measurements 
of $\ATodd$ in charm meson decays~\cite{HFLAV:2019otj,Belle:2022gjv,Belle:Moon}. 
Our results are the first such measurements for these decay modes and are among the world's most precise measurements for charm decays.

We also measure $\ATodd$ in subregions of phase space corresponding to intermediate 
processes $D^+\to\phi\rho^+$, $\Kstarzb K^{*+}$, and $\Kstarzb\rho^+$; and 
$D_s^+\to K^{*+}\rho^{0}$, $K^{*0}\rho^{+}$, $\phi\rho^+$, and $\Kstarzb K^{*+}$.
These results, as listed in Tab.~\ref{tab:ATodd4DToVV},
are also consistent with zero and show no evidence of 
$\CP$ violation.

\section*{Acknowledgments}
This work, based on data collected using the Belle detector, which was
operated until June 2010, was supported by 
the Ministry of Education, Culture, Sports, Science, and
Technology (MEXT) of Japan, the Japan Society for the 
Promotion of Science (JSPS), and the Tau-Lepton Physics 
Research Center of Nagoya University; 
the Australian Research Council including grants
DP210101900, 
DP210102831, 
DE220100462, 
LE210100098, 
LE230100085; 
Austrian Federal Ministry of Education, Science and Research (FWF) and
FWF Austrian Science Fund No.~P~31361-N36;
National Key R\&D Program of China under Contract No.~2022YFA1601903,
National Natural Science Foundation of China and research grants
No.~11575017,
No.~11761141009, 
No.~11705209, 
No.~11975076, 
No.~12135005, 
No.~12150004, 
No.~12161141008, 
and
No.~12175041, 
and Shandong Provincial Natural Science Foundation Project ZR2022JQ02;
the Ministry of Education, Youth and Sports of the Czech
Republic under Contract No.~LTT17020;
the Czech Science Foundation Grant No. 22-18469S;
Horizon 2020 ERC Advanced Grant No.~884719 and ERC Starting Grant No.~947006 ``InterLeptons'' (European Union);
the Carl Zeiss Foundation, the Deutsche Forschungsgemeinschaft, the
Excellence Cluster Universe, and the VolkswagenStiftung;
the Department of Atomic Energy (Project Identification No. RTI 4002) and the Department of Science and Technology of India; 
the Istituto Nazionale di Fisica Nucleare of Italy; 
National Research Foundation (NRF) of Korea Grant
Nos.~2016R1\-D1A1B\-02012900, 2018R1\-A2B\-3003643,
2018R1\-A6A1A\-06024970, RS\-2022\-00197659,
2019R1\-I1A3A\-01058933, 2021R1\-A6A1A\-03043957,
2021R1\-F1A\-1060423, 2021R1\-F1A\-1064008, 2022R1\-A2C\-1003993;
Radiation Science Research Institute, Foreign Large-size Research Facility Application Supporting project, the Global Science Experimental Data Hub Center of the Korea Institute of Science and Technology Information and KREONET/GLORIAD;
the Polish Ministry of Science and Higher Education and 
the National Science Center;
the Ministry of Science and Higher Education of the Russian Federation, Agreement 14.W03.31.0026, 
and the HSE University Basic Research Program, Moscow; 
University of Tabuk research grants
S-1440-0321, S-0256-1438, and S-0280-1439 (Saudi Arabia);
the Slovenian Research Agency Grant Nos. J1-9124 and P1-0135;
Ikerbasque, Basque Foundation for Science, Spain;
the Swiss National Science Foundation; 
the Ministry of Education and the Ministry of Science and Technology of Taiwan;
and the United States Department of Energy and the National Science Foundation.
These acknowledgements are not to be interpreted as an endorsement of any
statement made by any of our institutes, funding agencies, governments, or
their representatives.
We thank the KEKB group for the excellent operation of the
accelerator; the KEK cryogenics group for the efficient
operation of the solenoid; and the KEK computer group and the Pacific Northwest National
Laboratory (PNNL) Environmental Molecular Sciences Laboratory (EMSL)
computing group for strong computing support; and the National
Institute of Informatics, and Science Information NETwork 6 (SINET6) for
valuable network support.

\bibliography{references.bib}

\end{document}